\documentclass{article}

\usepackage{PRIMEarxiv}

\usepackage[utf8]{inputenc} 
\usepackage[T1]{fontenc}    
\usepackage{hyperref}       
\usepackage{url}            
\usepackage{booktabs}       
\usepackage{amsfonts}       
\usepackage{nicefrac}       
\usepackage{microtype}      
\usepackage{lipsum}
\usepackage{fancyhdr}       
\usepackage{graphicx}       
\graphicspath{{media/}}     
\usepackage{amsmath} 
\usepackage{threeparttable}
\usepackage{array}
\usepackage{footmisc,hyperref}

\usepackage{authblk}

\title{EEG-DCNet: A Fast and Accurate MI-EEG Dilated CNN Classification Method}

\author[1]{Wei Peng\#}
\author[1]{Kang Liu\#}
\author[2]{Jiaxi Shi}
\author[1,2]{Jianchen Hu*}

\affil[1]{School of the Future Technology, Xi'an Jiaotong University, Xi'an, China}
\affil[2]{School of the Automation Science and Engineering, Xi'an Jiaotong University, Xi'an, China}

\begin{document}
\maketitle

\footnotetext[1]{These \# authors contributed equally to this work.}
\footnotetext[2]{horace89@xjtu.edu.cn}

\begin{abstract}
The electroencephalography (EEG)-based motor imagery (MI) classification is a critical and challenging task in brain-computer interface (BCI) technology, which plays a significant role in assisting patients with functional impairments to regain mobility. We present a novel multi-scale atrous convolutional neural network (CNN) model called EEG-dilated convolution network (DCNet) to enhance the accuracy and efficiency of the EEG-based MI classification tasks. We incorporate the $1\times1$ convolutional layer and utilize the multi-branch parallel atrous convolutional architecture in EEG-DCNet to capture the highly nonlinear characteristics and multi-scale features of the EEG signals. Moreover, we utilize the sliding window to enhance the temporal consistency and utilize the  attension mechanism to improve the accuracy of recognizing user intentions. The experimental results (via the BCI-IV-2a ,BCI-IV-2b and the High-Gamma datasets) show that EEG-DCNet outperforms existing state-of-the-art (SOTA) approaches in terms of classification accuracy and Kappa scores. Furthermore, since EEG-DCNet requires less number of parameters, the training efficiency and memory consumption are also improved. The experiment code is open-sourced at \href{https://github.com/Kanyooo/EEG-DCNet}{here}.
\end{abstract}


\section{Introduction}\label{sec_intro}
Brain-computer interface (BCI), as one of the most well known technologies to enable individuals to control external devices with their own limbs, has attracted lots of attentions. It can help patients with functional impairments regain mobility~\cite{wang2023eeg}. It is recognised that the most chanllenging task is to accurately identify the control signal from the brain of the patient via limbs. As an efficient signal obtained from non-invasive technique, EEG signal has been utilized in widespread due to its high temporal resolution~\cite{wu2024real}. Obviously, precisely deciphering the relationship between EEG signals and motor imagery (MI) is a crucial pathway for BCI~\cite{ang2012filter}. Unfortunately, EEG signals usually suffer from low signal-to-noise ratio, poor spatial resolution, and significant individual variability. Additionally, the underlying mechanisms are not fully understood, and there is a lack of effective feature extraction methods, posing significant challenges for accurate classification~\cite{alotaiby2015review}.

Since the EEG signals are collected from multiple in position electrodes on the scalp which can reflect the activity of different brain regions, these signals usually present particular spatial characteristics. Thus, the deep learning method has been applied in EEG signal classification problems due to its remarkble ability to learn complex spatial features from raw data \cite{badr2024review}. In recent years, a great number of Convolutional Neural Network (CNN)-based methods have appeared to significantly improve the prediction accuracy and efficiency of the EEG signal classification \cite{zhao2019multi,song2022eeg,altaheri2022physics,gao2019eeg,zhang2021sparsedgcnn}.

The foundamental reason is that CNN is good at processing high-dimensional data with spatial local correlations, extracting features along the spatial dimensions through convolutional kernels to capture spatial relationships in EEG signals~\cite{cong2023review}. Moreover, by using one-dimensional convolutions or adding temporal convolutional layers, CNNs can also capture the temporal dependencies, e.g., \cite{zhao2019multi} introduced a new three-dimensional EEG representation method, converting EEG signals into 3D arrays to retain spatial distribution information. On the other hand, the demanding real-time processing requirement in BCI systems is suitable for CNN model since most of the calculations can be performed in off-line parallel training stage \cite{krizhevsky2014one}. Another particular feature in EEG signals is that they are often affected by the subject's emotions and other physiological states, potentially introducing noise. Thus, the noise reduction procedure such as the pooling \cite{huber2021random,cong2023review} is necessary to avoid its influence on the feature extraction.

After the preprocessing, the EEG signals can be converted into two-dimensional representations such as time-frequency maps and power spectral density plots. Therefore, the image recognition techniques can be utilized in EEG signals classification. The work of  \cite{zhao2019multi} designed a multi-branch 3D CNN architecture, including networks with different receptive field sizes to extract features at different scales. They used a cropped training strategy to increase training samples and improve the model's generalization ability. This method demonstrated excellent performance and robustness across different subjects and significantly improved practicality even with only 9 sampling electrodes. The work of \cite{eldele2021attention} designed a class-aware loss function to effectively handle data imbalance issues, introduced a multi-resolution CNN and an attention feature reweighting module to extract features from different frequency bands and enhance feature learning.

Due to the temporal sequence characteristics of EEG signals, another alternative approach is to utilize the Recurrent Neural Networks (RNNs) and their variants Long Short-Term Memory (LSTM) networks \cite{shashidhar2022eeg,dutta2019multi,nikoupour2024robust,amin2021attention}. However, in current MI classification research, temporal sequence models are usually combined with other methods to jointly extract spatial and temporal features to improve the performance. In EEG signal classification, temporal networks are often combined with CNNs~\cite{amin2021attention}. In \cite{amin2021attention}, LSTM was used to construct a recurrent attention network to capture temporal dependencies between different EEG time segments, allowing the model to focus on periods when subjects were paying attention during the experiment while ignoring other periods. The work of \cite{soleymani2015analysis} used the LSTM-RNN and continuous conditional random fields for automatic and continuous emotion detection. The study also explored the interference effect of facial muscle activity on EEG signals and analyzed the supplementary information provided by EEG signals in the presence of facial expressions. In \cite{luo2018exploring}, LSTM and gate recurrent unit (GRU) were used to build deep RNN architectures to extract spatial-frequency-sequential relationships in EEG signals, improving performance in MI classification. 

Recently, various attention mechanisms have been applied in EEG classification networks~\cite{song2022eeg,altaheri2022physics,eldele2021attention,amin2021attention,li2020multi}. The attention mechanism was initially proposed to address the information bottleneck in sequence-to-sequence (Seq2Seq) models when processing long sequences \cite{vaswani2017attention}. Its core idea is to allow the model to selectively focus on different parts of the input sequence when processing the current step, rather than relying on a fixed hidden state. Applying the attention mechanisms to EEG signal processing can enable the model to focus on important temporal segments and spatial locations, improving the effectiveness of feature extraction by automatically ignoring irrelevant or noisy signal parts. In \cite{amin2021attention}, the attention mechanism was used to focus on features most relevant to MI tasks in EEG signals, combining attention-guided inception CNN and LSTM to adaptively focus on different spatial contexts and reduce overfitting issues through a two-layer attention mechanism. The work in \cite{li2020multi} proposed a multi-scale attention-based fusion CNN, which extracts spatiotemporal multi-scale features of EEG signals through spatial multi-scale modules, attention modules, temporal multi-scale modules, and dense fusion modules. The attention mechanism enhances the network's sensitivity to EEG signal features, allowing the network to selectively amplify valuable feature channels and suppress useless ones. The work of \cite{bagchi2022eeg} proposed an EEG-ConvTransformer network based on multi-head self-attention and temporal convolution. This network combines self-attention modules and convolutional filters to capture inter-regional interaction patterns and temporal patterns within a single module. The attention mechanism is used to capture inter-regional interaction patterns in EEG signals and learn the temporal patterns of the signals.

The work of \cite{huang2022eeg} developed a new deep learning model that enhances adaptability and robustness across different individuals by introducing the local reparameterization trick into CNNs. This method performs well in handling inter-individual variability but still faces certain challenges. The work of \cite{an2023dual} proposed a novel dual attention relation network, including temporal attention and aggregation attention modules, along with a fine-tuning strategy to adapt to the data distribution of unseen subjects. The attention mechanism is used to improve classification accuracy of EEG signals from unseen subjects, especially in cases of scarce labeled data. Compared with existing few-shot learning methods, this method exhibits superior performance. The work of \cite{altaheri2022physics} proposed an Attention-based Temporal Convolutional Network (ATCNet), combining scientific machine learning (SciML) with attention mechanisms and temporal convolutional network (TCN) architectures to improve the decoding performance of MI-EEG signals, representing one of the advanced approaches currently available.

In summary, current research on MI classification of EEG signals mainly builds upon CNN architectures, integrating temporal neural network architectures and attention mechanisms to extract and utilize key features. We propose a new CNN-based model which can achieve the highest classification accuracy while mantaining a high training efficiency on multiple datasets including BCI-IV-2a ,BCI-IV-2b and the High-Gamma datasets. The main contributions are summarized as follows,

\begin{itemize}
    \item we insert a specifically designed multi-scale atrous CNN block into the EEG-DCNet backbone to replace the average pooling so that it can enhance the feature representation capabilities and reduce the information transmission loss;

    \item we incorporate the efficient $1\times1$ convolutional layer and a multi-branch parallel atrous convolutional architecture, along with a sliding window and attention mechanisms in EEG-DCNet, which effectively integrates multi-scale information and captures continuous changes in EEG signals, thereby improving the classification accuracy;

    \item we transform the EEG time serise signal into image-like tensor, which can not only make EEG-DCNet maintain high accuracy but also enhance its generalization capabilities across multiple datasets so that it can fit for the most mainstream CNN-based networks.

\end{itemize}

The structure of this paper is as follows. Section~\ref{sec_intro} reviews the more advanced related work in recent years and introduces the main contributions of the paper. Section~\ref{sec_method} introduces the composition of the EEG-DCNet model and the design ideas of each module. Section~\ref{sec_result} presents the testing environment and results of the model. Section~\ref{sec_conclusion} summarizes the work of this paper and provides a summary of potential issues and future research directions.

\section{EEG-DCNet Architecture}\label{sec_method}

\begin{figure}
    \centering
    \includegraphics[width=1\linewidth]{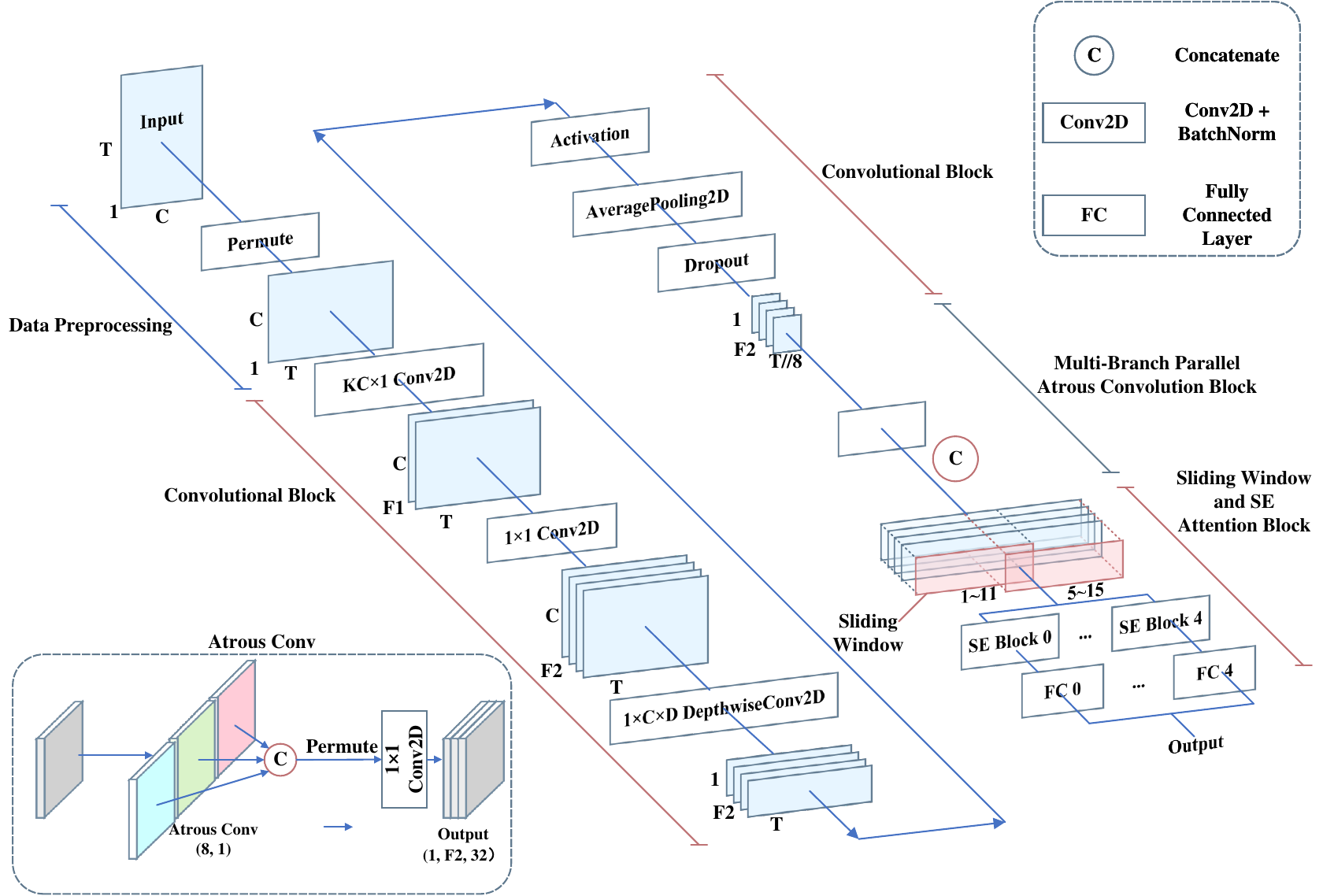}
    \caption{The architecture of EEG-DCNet. The architecture of EEG-DCNet consists of several key components: initial data preprocessing is followed by a convolutional block that extracts spatial and frequency features. An average pooling and dropout layer then perform dimensionality reduction and regularization. Next, a multi-branch parallel atrous convolution block captures multi-scale features at varying dilation rates. This is followed by a sliding window and SE attention module to enhance local feature representations. Finally, fully connected layers produce the output.}
    \label{fig_DCNet}
\end{figure}

In this section, we propose the EEG-DCNet, designed to enhance feature representation capabilities in EEG-based MI tasks. EEG-DCNet improves upon traditional linear two-dimensional (2D) convolutions and depthwise convolutions by incorporating \(1 \times 1\) convolutional layers, which enhance the network's ability to capture nonlinear characteristics of the data. Additionally, it introduces a multi-branch parallel atrous convolutional architecture to improve the perception of multi-scale features. Specifically, the \(1 \times 1\) convolutions are utilized to fuse outputs from atrous convolutions with different dilation rates, facilitating effective integration of multi-scale information. Then, a sliding window mechanism is employed to process the time-series EEG data, and when combined with an attention mechanism, it enables the network to capture continuous changes in EEG signals during the experiment, thereby enhancing temporal consistency and improving the accuracy in recognizing user intentions. The complete architecture of EEG-DCNet is depicted in Fig.~\ref{fig_DCNet}.

The input to EEG-DCNet consists of raw EEG signals with dimensions \( C \times T \), where \( C \) denotes the number of channels and \( T \) denotes the number of time samples. The signals are expanded by one dimension to serve as the input channel for convolutional operations. The output of the model is the predicted probability distribution over the four MI task categories. As illustrated in Fig.~\ref{fig_DCNet}, the EEG-DCNet architecture comprises three core modules: the convolutional (CV) module, the multi-branch parallel atrous convolution (SP) module, and the attention (AT) module.

\begin{itemize}
    \item \textbf{CV module:} By incorporating \(1 \times 1\) convolutions and applying nonlinear activation functions like ELU after each convolution operation, the network overcomes the limitations of linear operations in traditional 2D and depthwise convolutions, thereby enhancing its ability to capture the nonlinear characteristics inherent in EEG data.
    \item \textbf{SP module:} By incorporating multi-branch parallel atrous convolutions with varying dilation rates and fusing their outputs using a \(1 \times 1\) convolution, the network effectively captures multi-scale information, enhancing its ability to perceive complex features and ensuring that features extracted at different scales complement each other.
    \item \textbf{Sliding Window and AT modules:} By implementing a sliding window strategy to segment the time series into multiple local sub-sequences and integrating a channel attention mechanism (specifically, a Squeeze-and-Excitation block) within each window, the network increases training data, enhances temporal modeling precision, and selectively emphasizes important channels, thereby improving feature representation and ultimately boosting the accuracy of temporal predictions in recognizing user intentions.
\end{itemize}
Next, the design ideas of each module will be introduced one by one according to the forward order of DCNet.

\subsection{Data Preprocessing}

We train and fine-tune the architecture and parameters of EEG-DCNet on the BCI-2a dataset and test the performance on BCI-2a, BCI-2b and HGD datasets.
To preserve the integrity of the MI EEG signals, no preprocessing is applied to the raw EEG signals in this work. The input to the EEG-DCNet model consists of a raw EEG signal matrix with dimensions \( C \times T \), where \( C \) represents the number of electrode channels and \( T \) represents the number of time samples. The objective of EEG-DCNet is to map the input EEG signals to their corresponding class label \( y_i \), where \( y_i \in \{1, \dots, n\} \) is the class label for trial \( X_i \), and \( n \) is the total number of classes.

For the BCI-2a dataset, the specific input configuration is as follows: the number of time samples is \( T = 1125 \), the number of electrode channels is \( C = 22 \), and the classification task involves \( n = 4 \) MIclasses. The dataset contains \( m = 5184 \) MI trials.

\subsection{Convolutional Block}

\begin{figure}
    \centering
    \includegraphics[width=1\linewidth]{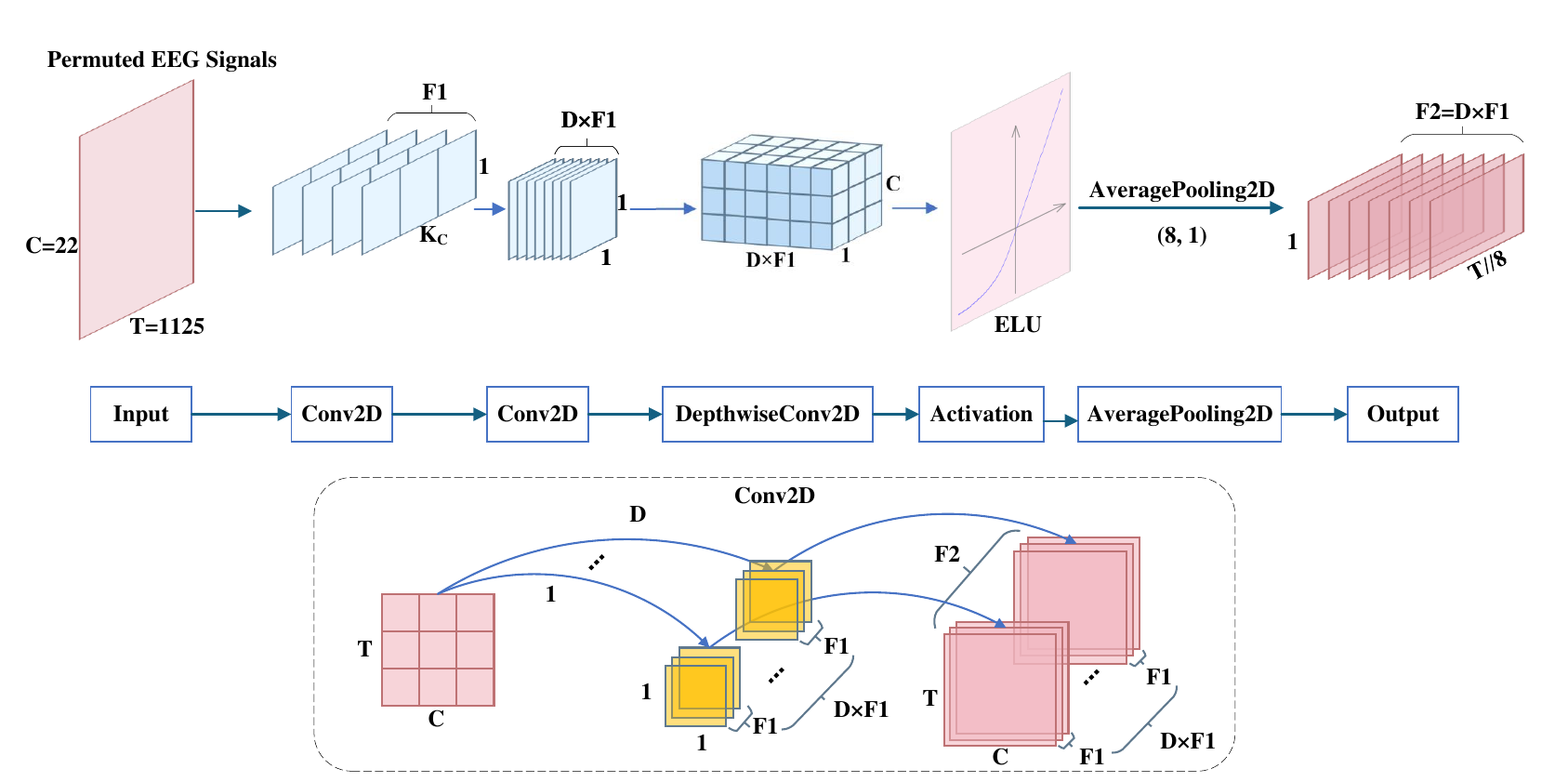}
    \caption{The architecture of CV block.}
    \label{fig_CV}
\end{figure}

The convolutional block consists of three convolutional layers, with a architecture similar to EEGNet. However, in contrast to EEGNet, EEG-DCNet introduces a $1\times1$ convolution layer in the depthwise convolution stage for dimensionality reduction. The detailed architecture is shown in Fig.~\ref{fig_CV} and the shape parameters of each layer are given in Table.~\ref{tab_CV}.

\begin{table}[h!]\label{tab_CV}
\centering
\begin{tabular}{@{}ccccccc@{}}
\toprule
\textbf{Layer}         & \textbf{\# Filters}     & \textbf{Size} & \textbf{\# Params} & \textbf{Output}       & \textbf{Activation} & \textbf{Options}           \\ \midrule
Input                  &                         &               &                    & (1, C, T)             &                     &                            \\
Permute                &                         &               &                    & (T, C, 1)             &                     &                            \\
Conv2D                 & F1                      & (KC, 1)       & KC $\times$ F1     & (T, C, F1)            &                     & mode = same                \\
BatchNorm              &                         &               & 2 $\times$ F1      & (T, C, F1)            &                     &                            \\
Conv2D                 & D $\times$ F1           & (1, 1)        & D $\times$ F1      & (T, C, F2)            &                     & mode = same                \\
BatchNorm              &                         &               & 2 $\times$ F2 & (T, C, F2)       &                     &                            \\
DepthwiseConv2D        & D $\times$ F1           & (1, C)        & 2 $\times$ D $\times$ F2 & (T, 1, F2$\times$ D)       &                     & mode = valid, depth = D \\
BatchNorm              &                         &               & 2 $\times$ D $\times$ F2     & (T, 1, F2$\times$ D)            &                     &                            \\
Activation             &                         &               &                    & (T, 1, F2$\times$ D)            & ELU                 &                            \\
AveragePooling2D       &                         & (8, 1)        &                    & (T//8, 1, F2)         &                     &                            \\
Dropout                &                         &               &                    & (T//8, 1, F2)         &                     & p = 0.25                   \\ \bottomrule
\end{tabular}
\caption{The convolutional block of EEG-DCNet}
\end{table}

The first layer applies \( F_1 \) filters of size \( (1, KC) \) to the input EEG signals, where \( KC \) is the temporal kernel length set to one-quarter of the sampling rate (for the BCI-2a dataset, \( KC = 64 \)). This layer performs temporal convolution along the time axis to extract multi-band temporal features, capturing frequency-specific patterns essential for distinguishing different MI tasks, and generates \( F_1 \) temporal feature maps representing the temporal dynamics of the EEG signals. 

Then, a \( 1 \times 1 \) convolution with \( F_2 \) filters is applied to the output of the first layer. This operation increases the number of feature maps, effectively expanding the feature dimensions and enhancing the network's capacity to represent complex patterns by allowing more combinations of features. As a result, it produces \( F_2 \) enhanced temporal feature maps with increased depth. 

Subsequently, the network employs depthwise convolution using \( F_2 \) filters of size \( (C, 1) \), where \( C \) represents the number of EEG channels. This layer captures spatial features by convolving across the electrode channels, modeling spatial dependencies and interactions between different scalp regions. It results in \( F_1 \times D \) feature maps, where \( D \) is the number of depthwise filters per input channel (empirically set to 2), effectively learning spatial filters specific to each temporal feature map. 

To reduce the dimensionality of the feature maps and decrease computational complexity while retaining essential information, a \( 1 \times 1 \) convolution layer is introduced after the depthwise convolution layer, producing a compact set of feature maps suitable for subsequent processing. A nonlinear activation function, such as the Exponential Linear Unit (ELU, the formulation is shown as ~\eqref{eq_elu}), is applied after each convolutional operation to introduce nonlinearity. This allows the network to model complex, nonlinear relationships inherent in EEG data that linear operations cannot capture, thereby enhancing the network's representational capacity and improving its ability to distinguish between different EEG signal patterns associated with various MI tasks. 
\begin{equation}\label{eq_elu}
\text{ELU}(x) = 
\begin{cases} 
x & \text{if } x > 0 \\
\alpha (e^{x} - 1) & \text{if } x \leq 0 
\end{cases}
\end{equation}
Finally, an average pooling layer of size \( (8, 1) \) is applied to the output feature maps to reduce the temporal dimension, downsampling the data to approximately 32 Hz. This step decreases computational complexity and focuses on the most salient temporal features, providing condensed feature representations that retain critical information for classification. Through this series of operations, the convolutional module effectively extracts and refines both temporal and spatial features from the EEG signals, laying a robust foundation for the subsequent modules in EEG-DCNet.

\subsection{Multi-Branch Parallel Atrous Convolution Block}

\begin{figure}
    \centering
    \includegraphics[width=0.7\linewidth]{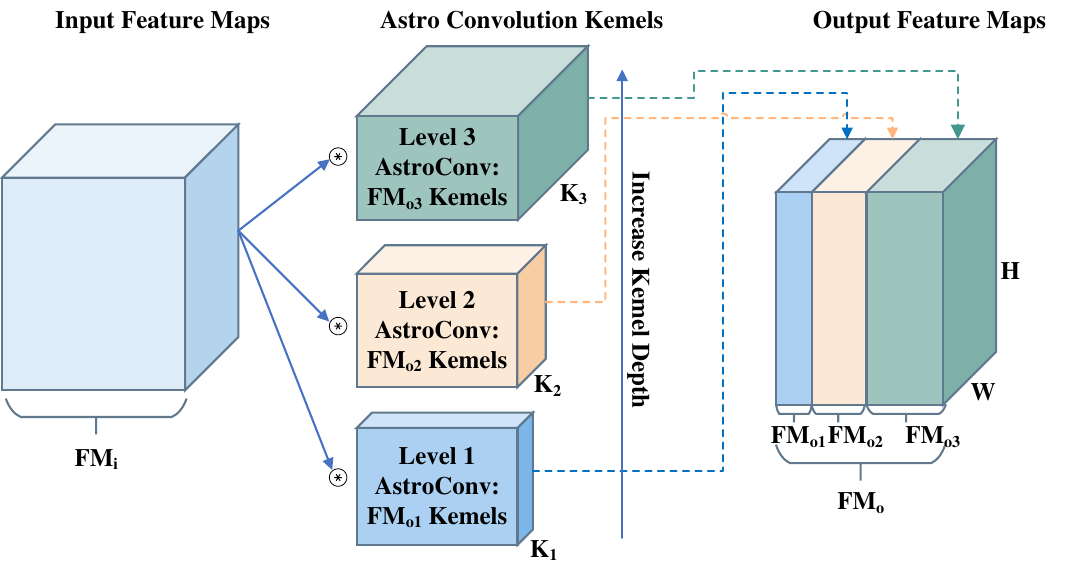}
    \caption{The architecture of AsrtoConv.}
    \label{fig_AsrtoConv}
\end{figure}

After the convolutional block, we incorporate a multi-branch parallel atrous convolution architecture to process the EEG feature maps at different scales. This architecture combines \( 1 \times 1 \) convolutions and parallel atrous convolutions, where the dilation rate of each atrous convolution is customized to flexibly enhance the model's ability to capture multi-scale features. Drawing inspiration from the atrous spatial pyramid pooling (ASPP) method, we design a parallel atrous convolution layer that uses multi-scale receptive fields, processing and integrating features from different scales independently across multiple branches. The architecture of SP block is shown as Fig.~\ref{fig_AsrtoConv} and the shape parameters are given in Table~\ref{tab_SP}.
\begin{table}[h!]\label{tab_SP}
\centering
\begin{tabular}{>{\raggedright\arraybackslash}p{2cm} >{\centering\arraybackslash}p{1.5cm} >{\centering\arraybackslash}p{1.5cm} >{\centering\arraybackslash}p{1.5cm} >{\centering\arraybackslash}p{2.5cm} >{\centering\arraybackslash}p{1.5cm} >{\centering\arraybackslash}p{2.5cm}}
\toprule
\textbf{Layer} & \textbf{\# Filters} & \textbf{Size} & \textbf{\# Params} & \textbf{Output} & \textbf{Activation} & \textbf{Options} \\
\midrule
Input          & -            & -          & -              & (T//8, 1, F2 * D)      & - & - \\
Atrous Conv1   & F2           & (8, 1)     & 8$\times$F2//4             & (T//8, 1, F2)         & - & dilation\_rate=2, use\_bias=False \\
BatchNorm      & -            & -          & 8$\times$F2              & (T//8, 1, F2)         & - & - \\
Activation     & -            & -          & -              & (T//8, 1, F2)         & ELU & - \\
Dropout        & -            & -          & -              & (T//8, 1, F2)         & - & p=0.25 \\
Atrous Conv2   & F2           & (8, 1)     & 8$\times$F2//4               & (T//8, 1, F2)         & - & dilation\_rate=4, use\_bias=False \\
BatchNorm      & -            & -          & 8$\times$F2             & (T//8, 1, F2)         & - & - \\
Activation     & -            & -          & -              & (T//8, 1, F2)         & ELU & - \\
Dropout        & -            & -          & -              & (T//8, 1, F2)         & - & p=0.25 \\
Atrous Conv3   & F2           & (8, 1)     & 8$\times$F2//4              & (T//8, 1, F2)         & - & dilation\_rate=6, use\_bias=False \\
BatchNorm      & -            & -          & 8$\times$F2              & (T//8, 1, F2)         & - & - \\
Activation     & -            & -          & -              & (T//8, 1, F2)         & ELU & - \\
Dropout        & -            & -          & -              & (T//8, 1, F2)         & - & p=0.25 \\
Concatenate    & -            & -          & -              & (T * 3//8, 1, F2)     & - & - \\
Permute        & -            & -          & -              & (1, F2, T * 3//8)     & - & - \\
Conv2D         & 32           & (1, 1)     & -              & (1, F2, 32)          & - & - \\
BatchNorm      & -            & -          & 2$\times$32             & (1, F2, 32)          & - & - \\
Activation     & -            & -          & -              & (1, F2, 32)          & ELU & - \\
Dropout        & -            & -          & -              & (1, F2, 32)          & - & p=0.25 \\
\bottomrule
\end{tabular}
\caption{Layer configuration of EEG-DCNet.}
\end{table}

For an atrous convolution with dilation rate \( r \) and kernel size \( k \), the receptive field is calculated as

\begin{equation}\label{eq_MBPAC}
    \text{Receptive Field} = 2 \times (r - 1) \times (k - 1) + k
\end{equation}

In this work, the multi-branch parallel atrous convolution architecture consists of three branches, each with a different dilation rate \( R = (2, 4, 6) \) and a kernel size of \( 8 \times 1 \). The receptive fields of these branches are \( 17 \times 1 \), \( 50 \times 1 \), and \( 78 \times 1 \), respectively. 
By using this multi-branch design, the model can efficiently capture features at different scales, thereby improving its ability to represent complex EEG signals and providing a richer set of features for subsequent layers. This approach significantly enhances the network's capability to process multi-scale information, which is crucial for interpreting the diverse temporal patterns in EEG data.

As shown in the Fig~\ref{fig_AsrtoConv}, the input consists of raw feature maps, which are processed by atrous convolutions with different dilation rates in parallel branches. These features are then concatenated along the channel dimension. To reduce the computational cost, a \( 1 \times 1 \) convolution is applied before concatenation to reduce the number of channels, thus optimizing resource utilization.

\subsection{Sliding Window and SE Attention Block}

\begin{figure}
    \centering
    \includegraphics[width=0.9\linewidth]{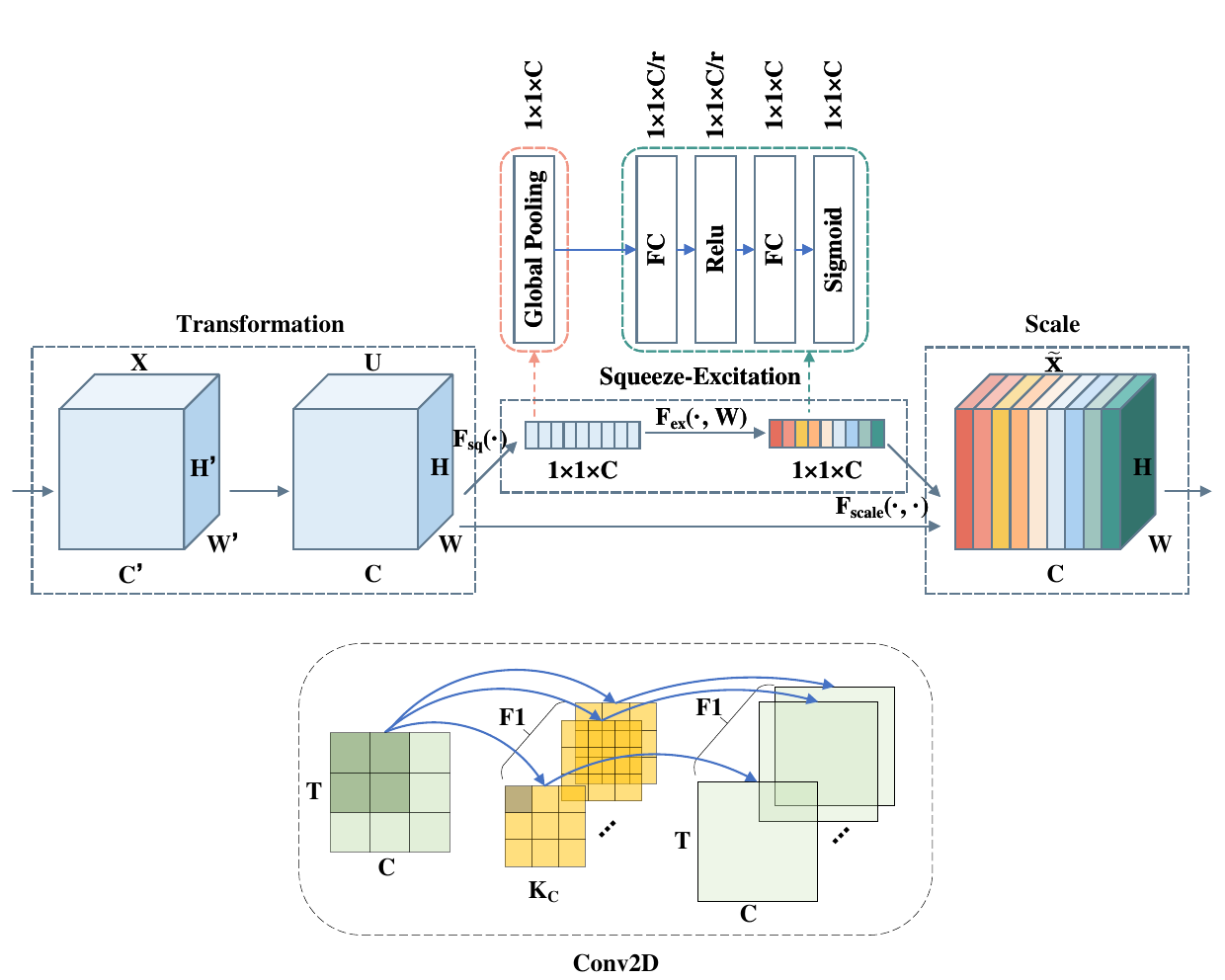}
    \caption{The architecture of EEG-DCNet.}
    \label{fig_SWSE}
\end{figure}

After processing the EEG signals through the CV and SP modules to capture frequency domain features, the sliding window and attention mechanism will be applied to extract the temporal domain information. The sliding window method helps expand the data volume and improve classification accuracy. We divide the time series into multiple overlapping windows of length \( T_w \), and each window \( z_i^w \) is fed into the attention block (AT) for classification prediction. Subsequently, the SoftMax classifier calculates the probability for each class, and the results are passed to the fully connected layer (FC), where the average prediction of all windows is computed to obtain the final classification output. The architecture of SP block is shown as Fig.~\ref{fig_SWSE} and the shape parameters are given in Table~\ref{tab_SWSE}.

\begin{table}[h!]
\centering
\begin{tabular}{>{\raggedright\arraybackslash}p{2.5cm} >{\centering\arraybackslash}p{1.5cm} >{\centering\arraybackslash}p{2cm} >{\centering\arraybackslash}p{1.5cm} >{\centering\arraybackslash}p{5cm}}
\toprule
\textbf{Layer} & \textbf{\# Params} & \textbf{Output} & \textbf{Activation} & \textbf{Options} \\
\midrule
Input          & -              & (1, F2, 32)      & -             & - \\
Lambda         & -              & (F2, 32)         & -             & Lambda(lambda x: x[:, -1, :, :]) \\
Permute        & -              & (32, F2)         & -             & - \\
Sliding windows & -              & (27, F2)         & -             & 1–27, 2–28, 3–29, 4–30, 5–31, 6–32 \\
SE block       & -              & (27, F2)         & -             & - \\
Lambda         & -              & F2          & -             & Lambda(lambda x: x[:, -1, :]) \\
Dense          & -              & n\_classes  & -             & - \\
Average        & -              & n\_classes                & -             & tf.keras.layers.Average()(sw\_concat[:]) \\
\bottomrule
\end{tabular}
\caption{Additional layer configuration of EEG-DCNet.}
\label{tab_SWSE}
\end{table}
The sliding window method enables effective capture of features over different time segments during a trial, rather than focusing on a single time point, which is important in MI tasks where the specific time points may vary across trials and sessions. In BCI tasks, the user may not immediately respond after receiving the cue, often exhibiting a response delay, and the response time can vary significantly between individuals. Analyzing the EEG signal over different time segments within a sliding window can help to obtain more consistent and reliable classification results. 

In this work, we adopt a convolutional-based sliding window approach, which integrates convolutional layers within each window processing step to efficiently capture localized temporal features. This approach enables parallel processing of all windows, thereby reducing training and inference time. The length of the sliding window \( T_w \) is calculated by the following formulas, depending on the pooling operations applied to the sequence

\begin{equation}
    T_w = T_c - n + 1, \quad T_c > n \geq 1
\end{equation}

\begin{equation}
    T_w = \frac{T}{8P_2} - n + 1
\end{equation}

where \( T_c \) represents the total length of the time series after pooling, \( T \) is the original sequence length, \( P_2 \) is the pooling size, and \( n \) represents the step size of the window. By applying these formulas, the sliding window method can dynamically adapt to different time resolutions, effectively balancing computational efficiency with temporal feature richness. This strategy enhances the model's ability to capture complex temporal dependencies in EEG signals, thus leading to improved classification performance across trials.

Attention mechanisms in neuroscience imitate the human brain's ability to selectively focus on important information while ignoring irrelevant information, thus enhancing the efficiency and accuracy of information processing. In deep neural networks, attention mechanisms work similarly by adaptively enhancing key feature representations while suppressing less important features, improving the model's performance.

The SE (Squeeze-and-Excitation) block is a channel attention mechanism that recalibrates channel responses. It first applies global average pooling to compress spatial information into a channel descriptor. A two-layer fully connected network then reduces the dimensionality and generates modulation weights using ReLU and sigmoid activations. These weights are applied to the original feature map, enhancing relevant features and suppressing irrelevant information. Fig.~\ref{fig_SWSE} show the architecture of the SE block and detailed calculation procedure is given below.

The whole calculation can be divided into four parts.
\begin{enumerate}
    \item\textbf{Input Transformation} \\
    The input feature map $X$ is mapped to $U \in \mathbb{R}^{H \times W \times C}$, where $H$ and $W$ represent spatial dimensions, and $C$ represents the number of channels. This transformation typically involves a convolution operation to adjust the dimensions of the input feature map.

    \item \textbf{Squeeze (Global Average Pooling)} \\
    Since convolution operations capture only local context, the squeeze operation uses global average pooling to compress the global spatial information into a channel descriptor. The transformed feature map $U$ is pooled across the spatial dimensions (height $H$ and width $W$), generating a channel-wise descriptor:
    \begin{equation}
    z = F_{sq}(U) = \frac{1}{H \times W} \sum_{i=1}^H \sum_{j=1}^W U_{i,j}
    \end{equation}
     where $z \in \mathbb{R}^{1 \times 1 \times C}$ represents the globally aggregated channel information.

    \item \textbf{Excitation (Two-layer Fully Connected Network)} \\
    Following the squeeze operation, the excitation operation utilizes a two-layer fully connected network to recalibrate the channel-wise feature responses. This process can be broken down as follows:
    The first fully connected layer reduces the dimensionality of the channel descriptor by a reduction ratio $r$ and applies the ReLU activation function to introduce non-linearity:
    \begin{equation}
    s = \text{ReLU}(W_1 z + b_1)
    \end{equation}
     where $W_1 \in \mathbb{R}^{C/r \times C}$ and $b_1 \in \mathbb{R}^{C/r}$ are the weights and biases for the first fully connected layer.
    The second fully connected layer restores the dimensionality from $C/r$ back to $C$ and applies a sigmoid activation function to produce a modulation weight for each channel:
    \begin{equation}
     s = \sigma(W_2 s + b_2)
    \end{equation}
     where $W_2 \in \mathbb{R}^{C \times C/r}$ and $b_2 \in \mathbb{R}^{C}$ are the weights and biases for the second fully connected layer, and $\sigma$ represents the sigmoid activation function. The output $s \in \mathbb{R}^{1 \times 1 \times C}$ provides the channel-wise modulation weights.

    \item \textbf{Scale} \\
    Finally, the scale operation applies these weights to the original feature map $U$, where the weights $s$ are multiplied with the corresponding channels of the feature map. This operation selectively enhances relevant features while suppressing irrelevant information:
    \begin{equation}
    \tilde{X} = F_{scale}(U, s) = U \odot s
    \end{equation}
     where $\odot$ denotes element-wise multiplication. Each channel in $U$ is scaled by the corresponding value in $s$, producing the recalibrated feature map $\tilde{X} \in \mathbb{R}^{H \times W \times C}$.
\end{enumerate}

\section{Experiment and Result Analysis}\label{sec_result}

The EEG-DCNet model, along with comparison models, was trained and tested using the TensorFlow framework on a single GPU (Nvidia GTX 3070Ti with 8 GB of memory). For all experiments, the following training configuration was applied:

The model was optimized using the Adam optimizer with a learning rate of $0.001$, a batch size of $64$, and trained for $500$ epochs, with early stopping set to a patience of $300$ epochs. The proposed EEG-DCNet model achieved an overall accuracy of $87.94\%$ and a Kappa score of $0.8392$, outperforming existing SOTA approaches on all test datasets. Additionally, the EEG-DCNet model requires fewer parameters and demonstrates faster training times compared to existing models, further highlighting its efficiency and effectiveness in EEG classification tasks.

\subsection{Dataset}
Our approach is evaluated on three widely used EEG datasets: the BCI Competition IV dataset 2a, the BCI Competition IV dataset 2b, and the High Gamma Dataset. These datasets were collected using different acquisition devices, experimental paradigms, numbers of subjects, and sample sizes, enabling us to assess the generalizability of our approach.

Dataset 1: The BCI-IV-2a dataset consists of EEG data from 9 subjects, each of whom performed four different MI tasks during the experiment: left hand (class 1), right hand (class 2), both feet (class 3), and tongue (class 4) MI. For each subject, two sessions were recorded on different days, one used for training (T) and the other for testing (E) the classifier. Each session was further divided into 6 runs, with each run containing 48 trials. Each trial was recorded using 25 channels (including 22 EEG channels and 3 EOG channels), with a sampling rate of 250 Hz and a bandpass filter from 0.5 Hz to 100 Hz.

Dataset 2: The BCI-IV-2b dataset consists of EEG data from 9 right-handed subjects with normal or corrected-to-normal vision. During the experiment, the subjects performed two different MI tasks: left hand (class 1) and right hand (class 2) MI. For each subject, EEG data were recorded over 5 sessions. The first two sessions involved MI without visual feedback, while the last three sessions included visual feedback. The non-feedback sessions were divided into 4 runs, and the feedback sessions into 2 runs, with each run consisting of 20 trials. Each trial was recorded from 3 channels (C3, Cz, C4), with a sampling rate of 250 Hz and a bandpass filter from 0.5 Hz to 100 Hz.

Dataset 3: The High Gamma Dataset consists of EEG data from 14 subjects, each performing four different MI tasks during the experiment: left hand (class 1), right hand (class 2), both feet (class 3), and rest (class 4) MI. For each subject, approximately 1000 four-second MI trials were conducted, divided into 13 runs. Each trial was recorded using 128 channels.

\subsection{Evaluation Metrics}
The proposed model in this paper is evaluated using accuracy and the Kappa score. The accuracy measures the proportion of correct predictions made by the model over all classes and the Kappa score evaluates the agreement between predicted and true labels, adjusting for the chance of agreement. Equations \eqref{eq_accuracy} and \eqref{eq_kappa} give their calculation methods respectively.

\begin{equation}\label{eq_accuracy}
\text{Accuracy} = \frac{\sum_{i=1}^{n} TP_i / I_i}{n}
\end{equation}
where $TP_i$ denotes the number of correctly predicted samples for class $i$, $I_i$ is the total number of samples in class $i$, and $n$ is the number of classes.

\begin{equation}\label{eq_kappa}
\text{Kappa} = \frac{1}{n} \sum_{a=1}^{n} \frac{P_a - P_e}{1 - P_e}
\end{equation}
where $P_a$ is the average accuracy across all classes, and $P_e$ is the expected accuracy by random guessing.

\subsection{Comparison of number of sliding windows }

Table~\ref{tab_Different_Windows} illustrates the relationship between the number of sliding windows and the accuracy of the EEG-DCNet model. Currently, there is no analytical method to determine the optimal number of sliding windows directly. Instead, it must be tested sequentially to identify the setting that yields the highest accuracy. The results show that increasing the number of sliding windows generally improves performance up to a certain point (specifically, $6$ windows in this case). However, further increases beyond this optimal number can lead to fluctuations in accuracy and Kappa scores, as observed in the table. This behavior suggests that while more windows may capture finer temporal features, excessive windows may introduce redundancy or noise, reducing model efficiency and stability.

\begin{table}[h!]\label{tab_Different_Windows}
\centering
\begin{tabular}{@{}ccccccccccccc@{}}
\toprule
\textbf{Windows} & \textbf{S01} & \textbf{S02} & \textbf{S03} & \textbf{S04} & \textbf{S05} & \textbf{S06} & \textbf{S07} & \textbf{S08} & \textbf{S09} & \textbf{Average} & \textbf{Kappa} \\ \midrule
1                & 87.67        & 80.03        & 93.92        & 86.11        & 87.15        & 78.30        & 75.52        & 89.93        & 88.89        & 85.28          & 0.7971          \\
2                & 87.15        & 81.77        & 95.49        & 81.94        & 88.72        & 80.90        & 82.47        & 80.62        & 89.76        & 86.54          & 0.8205         \\
3                & 89.93        & 82.29        & 94.97        & 84.55        & 86.63        & 83.85        & 73.78        & 91.49        & 91.15        & 86.52          & 0.8208         \\
4                & 89.06        & 83.68        & 93.92        & 83.51        & 88.02        & 81.08        & 80.90        & 90.80        & 82.01        & 87.00          & 0.8266         \\
5                & 88.02        & 83.51        & 89.76        & 84.38        & 90.62        & 80.90        & 87.85        & 92.88        & 88.02        & 87.33          & 0.8310         \\
6                & 90.45        & 85.07        & 93.40        & 84.55        & 88.37        & 79.34        & 88.37        & 91.32        & 90.62        & 87.94          & 0.8392         \\
7                & 90.28        & 79.34        & 95.83        & 89.06        & 88.54        & 82.12        & 63.72        & 90.62        & 90.62        & 85.57          & 0.8076         \\
8                & 90.28        & 79.86        & 95.31        & 86.63        & 86.63        & 82.12        & 69.79        & 91.32        & 92.01        & 86.00          & 0.8376          \\
9                & 91.32        & 83.51        & 97.22        & 85.24        & 88.72        & 80.38        & 61.46        & 91.84        & 90.80        & 85.61          & 0.8081         \\
10               & 90.45        & 85.07        & 95.83        & 85.94        & 89.76        & 82.64        & 78.99        & 91.49        & 90.10        & 87.81          & 0.8374         \\
11               & 92.88        & 80.21        & 97.22        & 87.67        & 89.76        & 81.77        & 62.15        & 92.19        & 89.93        & 85.98          & 0.8130         \\
12               & 93.92        & 83.68        & 98.09        & 88.37        & 88.37        & 82.64        & 77.78        & 92.71        & 90.28        & 88.43          & 0.8457         \\
13               & 88.19        & 87.50        & 97.05        & 85.94        & 87.67        & 82.81        & 74.13        & 91.32        & 90.45        & 87.23          & 0.8297         \\ \bottomrule
\end{tabular}
\caption{Performance Evaluation of Different Windows}
\end{table}

\subsection{Comparison of Different Attention Mechanisms}

Table~\ref{tab_Different_attentions} presents a comparison of four attention mechanisms: CBAM, CA, ECA, and SE—demonstrating their impact on the performance of the EEG-DCNet model. The results indicate that all attention mechanisms contribute to enhancing DCNet's performance, with CBAM and SE showing superior results in terms of accuracy and Kappa score.

However, CBAM operates significantly slower than SE. This suggests that SE may be more suitable for applications involving 2-D EEG features, because it can balance the accuracy and computational performance.

\begin{table}[h!]\label{tab_Different_attentions}
\centering
\begin{tabular}{@{}cccccccccccc@{}}
\toprule
\textbf{Attention} & \textbf{S01} & \textbf{S02} & \textbf{S03} & \textbf{S04} & \textbf{S05} & \textbf{S06} & \textbf{S07} & \textbf{S08} & \textbf{S09} & \textbf{Average} & \textbf{Kappa} \\ \midrule
None               & 90.28        & 80.56        & 94.62        & 86.11        & 88.89        & 76.39        & 84.55        & 90.80        & 85.07        & 86.36          & 0.8083          \\
CBAM               & 91.32        & 81.94        & 94.70        & 82.47        & 89.93        & 82.64        & 76.39        & 92.71        & 92.01        & 87.35          & 0.8313          \\
CA                 & 87.15        & 82.12        & 93.75        & 83.68        & 88.02        & 79.51        & 85.59        & 91.49        & 87.15        & 86.50          & 0.8200          \\
ECA                & 90.45        & 81.08        & 94.27        & 83.68        & 89.24        & 82.29        & 81.42        & 89.76        & 86.98        & 86.57          & 0.8209          \\
SE                 & 90.45        & 85.07        & 93.40        & 84.55        & 88.37        & 79.34        & 88.37        & 91.32        & 90.62        & 87.94          & 0.8392          \\ \bottomrule
\end{tabular}
\caption{Performance Comparison of Different Attention Mechanisms}
\end{table}

\subsection{Ablation Analysis}

In this part, we conduct an ablation analysis to evaluate the contribution of each block within the EEG-DCNet model to the overall performance on MI classification using the BCI-2a dataset. Table~\ref{tab_Different Block} shows the performance impact of each block in terms of accuracy and Kappa score. Each block was systematically removed from the model before training and validation to isolate its effect on classification performance.

The results indicate that incorporating the SP block increases the model’s accuracy by $4.84\%$, the SW block by $2.83\%$, and the AT block by $1.58\%$. These findings highlight that each block contributes a unique and valuable improvement to the model's performance, confirming that the combination of all blocks yields the highest accuracy and Kappa score. This analysis underscores the importance of each component within the EEG-DCNet architecture and demonstrates that their integration enhances classification performance on EEG data.

\begin{table}[h!]\label{tab_Different Block}
\centering
\begin{tabular}{@{}cccccccccccc@{}}
\toprule
\textbf{Block}        & \textbf{S01} & \textbf{S02} & \textbf{S03} & \textbf{S04} & \textbf{S05} & \textbf{S06} & \textbf{S07} & \textbf{S08} & \textbf{S09} & \textbf{Average} & \textbf{Kappa} \\ \midrule
None                  & 90.97        & 74.31        & 96.01        & 64.93        & 82.29        & 70.31        & 48.78        & 87.50        & 84.03        & 77.68          & 0.7024          \\
SP                    & 85.94        & 75.87        & 88.89        & 82.99        & 88.02        & 80.73        & 75.69        & 89.76        & 83.85        & 83.53          & 0.7562          \\
SP + SW               & 90.28        & 80.56        & 94.62        & 86.11        & 88.89        & 76.39        & 84.55        & 90.80        & 85.07        & 86.36          & 0.8083          \\
SP + SW + AT          & 90.45        & 85.07        & 93.40        & 84.55        & 88.37        & 79.34        & 88.37        & 91.32        & 90.62        & 87.94          & 0.8392          \\ \bottomrule
\end{tabular}
\caption{Performance Comparison of Different Block Configurations}
\end{table}

\subsection{Baseline Comparison}

This section presents a summary of the accuracy and Kappa scores of the proposed EEG-DCNet model based on the BCI-2a dataset, along with a comparison against several reproduced baseline models: EEGNet, ShallowConvNet, MBEEG\_SENet, EEGTCNet, EEGNeX, and ATCNet. For each of these models, the results are based on the hyperparameters specified in this study, with consistent preprocessing, training, and evaluation procedures applied across all models.
Below is a brief overview of the baseline models included in this comparison:

\begin{enumerate}
    \item \textbf{EEGNet}\cite{lawhern2018eegnet}: Utilizes 2D temporal convolutions, depthwise convolutions, and separable convolutions to provide consistent handling across various BCI tasks.
    \item \textbf{ShallowConvNet}\cite{wang2019shallow}: Uses two convolutional layers and one mean pooling layer to perform MI EEG classification.
    \item \textbf{MBEEG\_SENet}\cite{altuwaijri2022multi}: Employs a multi-branch CNN model with SE attention blocks to capture EEG features at varying time scales.
    \item \textbf{EEG-TCNet}\cite{ingolfsson2020eeg}: Integrates a CNN model with a TCN (Temporal Convolutional Network) module to capture temporal information in EEG signals.
    \item \textbf{EEGNeX}\cite{chen2024toward}: Applies two layers of 2D temporal convolutions and depthwise convolutions to extract both temporal and frequency information from EEG signals.
    \item \textbf{ATCNet}\cite{altaheri2022physics}: Implements MHA (Multi-Head Attention) with a sliding window approach, using a TCN module to enhance temporal feature extraction for EEG classification.
\end{enumerate}

As shown in Table~\ref{tab_result_2a}, EEG-DCNet achieves the highest performance among the models tested, with an average accuracy of $87.94\%$ and a Kappa score of $0.8392$. This represents a notable $10.26\%$ improvement in accuracy compared to the baseline EEGNet model.

\begin{table}[h!]
\label{tab_result_2a}
\centering
\begin{tabular}{@{}cccccccccccc@{}}
\toprule
\textbf{Methods}      & \textbf{S01} & \textbf{S02} & \textbf{S03} & \textbf{S04} & \textbf{S05} & \textbf{S06} & \textbf{S07} & \textbf{S08} & \textbf{S09} & \textbf{Average} & \textbf{Kappa} \\ \midrule
EEGNet                & 90.97        & 74.31        & 96.01        & 64.93        & 82.29        & 70.31        & 48.78        & 87.50        & 84.03        & 77.68          & 0.7024          \\
ShallowConvNet        & 83.68        & 75.52        & 92.19        & 64.24        & 81.77        & 72.92        & 78.30        & 92.36        & 83.68        & 80.52          & 0.7402          \\
MBEEG\_SENet          & 86.46        & 77.60        & 96.88        & 66.15        & 84.90        & 76.22        & 54.17        & 88.72        & 88.72        & 79.98          & 0.7330          \\
EEGTCNet              & 81.77        & 58.85        & 89.93        & 64.58        & 84.55        & 72.57        & 82.47        & 84.90        & 82.12        & 77.97          & 0.7063          \\
EEGNeX                & 85.07        & 82.64        & 83.85        & 82.64        & 89.76        & 79.86        & 85.59        & 83.68        & 82.99        & 84.01          & 0.7868          \\
ATCNet                & 92.01        & 79.34        & 97.74        & 82.12        & 87.67        & 85.59        & 53.12        & 91.32        & 94.27        & 84.80          & 0.7973          \\
EEG-DCNet                 & 90.45        & 85.07        & 93.40        & 84.55        & 88.37        & 79.34        & 88.37        & 91.32        & 90.62        & 87.94          & 0.8392          \\ \bottomrule
\end{tabular}
\caption{Performance Comparison of Different Methods on the BCI-2a Dataset}
\end{table}

Table~\ref{tab_result_2a} presents the performance comparison of various models on the BCI-2a dataset. The results show that EEG-DCNet outperforms all other models, achieving an average accuracy of $87.94\%$ and a Kappa score of $0.8392$. Compared to the baseline EEGNet model, EEG-DCNet demonstrates a $10.26\%$ improvement in accuracy, highlighting its enhanced ability to capture and interpret EEG features effectively. The ATCNet model is the closest competitor but falls short in both accuracy and Kappa score, underscoring DCNet's advantages in handling MI classification.

\begin{table}[ht!]
\label{tab_result_2b}
\centering
\begin{tabular}{@{}cccccccccccc@{}}
\toprule
\textbf{Methods}      & \textbf{S01} & \textbf{S02} & \textbf{S03} & \textbf{S04} & \textbf{S05} & \textbf{S06} & \textbf{S07} & \textbf{S08} & \textbf{S09} & \textbf{Average} & \textbf{Kappa} \\ \midrule
EEGNet                & 77.90        & 78.09        & 79.47        & 97.45        & 91.60        & 92.86        & 86.73        & 87.81        & 85.77        & 86.08          & 0.7213          \\
ShallowConvNet        & 80.62        & 79.27        & 73.00        & 99.15        & 93.74        & 92.12        & 84.01        & 84.05        & 89.86        & 86.20          & 0.7238          \\
MBEEG\_SENet          & 74.46        & 71.43        & 84.41        & 92.63        & 88.24        & 94.32        & 84.18        & 93.91        & 95.20        & 86.53          & 0.7302          \\
EEGTCNet              & 73.37        & 73.87        & 77.76        & 84.28        & 90.84        & 85.16        & 92.35        & 89.07        & 86.48        & 83.69          & 0.6731          \\
EEGNeX                & 73.25        & 75.51        & 88.26        & 96.09        & 97.07        & 82.47        & 92.24        & 95.22        & 81.22        & 86.81          & 0.7365          \\
ATCNet                & 81.34        & 67.25        & 85.55        & 98.58        & 98.63        & 92.49        & 93.88        & 94.09        & 92.88        & 89.41          & 0.7880          \\
EEG-DCNet                 & 88.41        & 86.93        & 89.16        & 98.58        & 96.18        & 93.96        & 91.50        & 95.34        & 91.81        & 92.43          & 0.8486          \\ \bottomrule
\end{tabular}
\caption{Performance Comparison of Different Methods on the BCI-2b Dataset}
\end{table}

Table~\ref{tab_result_2b} shows the model performances on the BCI-2b dataset, with EEG-DCNet again achieving the highest accuracy and Kappa score of $92.43\%$ and $0.8486$, respectively. EEG-DCNet consistently surpasses other models across individual subjects, particularly excelling over EEGNet and ShallowConvNet. These results emphasize DCNet's robust generalization across different datasets, reaffirming its efficacy for EEG classification tasks.

\begin{table}[ht!]
\label{tab_result_HGD}
\centering
\begin{tabular}{@{}ccc@{}}
\toprule
\textbf{Methods}      & \textbf{Average} & \textbf{Kappa} \\ \midrule
EEGNet                & 88.25            & 0.8433         \\
ShallowConvNet        & 87.00            & 0.8267         \\
MBEEG\_SENet          & 90.13            & 0.8684         \\
EEGTCNet              & 87.80            & 0.8373         \\
EEGNeX                & 87.58            & 0.8344         \\
ATCNet                & 92.05            & 0.8940         \\
EEG-DCNet             & 94.55            & 0.9273         \\ 
\bottomrule
\end{tabular}
\caption{Performance Comparison of Different Methods on the HGD Dataset}
\end{table}

In Table~\ref{tab_result_HGD}, EEG-DCNet achieves an average accuracy of $94.55\%$ and a Kappa score of $0.9273$ on the HGD dataset, outperforming the other methods tested. DCNet's significant margin over ATCNet, the second-best performing model, highlights its superior feature extraction and classification capabilities, making it especially well-suited for high-dimensional EEG data.

\begin{table}[ht!]\label{tab_param_compare}
\centering
\begin{tabular}{@{}cccc@{}}
\toprule
\textbf{Methods}    & \textbf{Mean Accuracy} & \textbf{FLOPs /M} & \textbf{Number of Parameters} \\ \midrule
EEGNet              & 77.68                  & 26.7               & 2.548 $\times$ 10\textsuperscript{3}     \\
ShallowConvNet      & 80.52                  & 127                & 47.31 $\times$ 10\textsuperscript{3}     \\
MBEEG\_SENet        & 79.98                  & 71.5               & 10.17 $\times$ 10\textsuperscript{3}     \\
EEGTCNet            & 77.97                  & 14.2               & 4.096 $\times$ 10\textsuperscript{3}     \\
EEGNeX              & 84.01                  & 444                & 63.626 $\times$ 10\textsuperscript{3}    \\
ATCNet              & 84.80                  & 60.5               & 113.732 $\times$ 10\textsuperscript{3}   \\
EEG-DCNet               & 87.94                  & 49                 & 28.64 $\times$ 10\textsuperscript{3}     \\ \bottomrule
\end{tabular}
\caption{Comparison of Methods in Terms of Mean Accuracy, FLOPs, and Number of Parameters}
\end{table}

In terms of computational efficiency, Table~\ref{tab_param_compare} shows that EEG-DCNet outperforms ATCNet with only one-quarter of the parameters, achieving $3.14\%$ higher accuracy. Furthermore, EEG-DCNet requires only one-tenth of the FLOPs of EEGNeX, yet achieves a $3.93\%$ increase in accuracy, highlighting its efficiency and effectiveness for EEG classification tasks. That means the EEG-DCNet is the SOTA model in EEG-MI classification task.

\section{Conclusion}\label{sec_conclusion}
The EEG-DCNet model proposed in this paper has demonstrated significant performance improvements in the EEG-MI classification task, validating its effectiveness in processing EEG signals and enhancing classification accuracy. By introducing 1×1 convolution layers and a multi-branch parallel dilated convolution structure, EEG-DCNet can capture the nonlinear features and multi-scale information of EEG signals, greatly improving the model's feature extraction capability. Additionally, by combining sliding windows and attention mechanisms, the model is able to better capture the temporal dynamics of EEG signals, thus enhancing the recognition of user intentions. Experimental results on three different EEG datasets show that EEG-DCNet outperforms existing SOTA approaches in terms of classification accuracy and Kappa score, while also demonstrating superior performance in terms of parameter count and computational efficiency. These results not only confirm the potential of EEG-DCNet as an efficient and accurate EEG-MI classification model, but also provide new directions for the future development of BCI technology.

However, EEG-DCNet still has many areas that need improvement, and there is still much to explore in the MI EEG classification task. Beyond the proposed blocks, we also tried adding temporal convolution modules after the sliding time window\cite{altaheri2022physics}, but the classification performance did not show evident improvement. The effectiveness of TCNs for EEG signals still requires further validation, especially in specific datasets and task scenarios, where TCNs may not necessarily outperform traditional methods. Furthermore, while EEG-DCNet's model architecture achieves the best performance on multiple datasets, it requires training from scratch each time, and its generalization ability across datasets is limited. This makes it challenging to perform real-time updates in practical applications, particularly when deployed on edge devices. Incremental learning is an effective approach to address this issue\cite{ma2022few}, and although there is limited research in this area, we have conducted several experiments and obtained promising results. The attention mechanism in the model helps in understanding key features of EEG signals. By analyzing the weights in the attention module, we can identify the EEG signal waveforms that play a crucial role in classification decisions, providing a better understanding of how the model makes decisions\cite{bastings2020elephant}. This is vital for enhancing clinical acceptance and user trust. Additionally, multimodal fusion offers a new opportunity for improving model performance. In recent years, classification using only EEG signals has gradually reached a performance ceiling, and combining other modalities, such as functional near infrared spectroscopy imaging data, can effectively reduce the impact of individual differences and help the model learn more common features, thereby improving classification accuracy.

\bibliographystyle{unsrt}  
\bibliography{references}

\end{document}